\documentclass{article}
\usepackage{spconf,amsmath,graphicx}
\usepackage{url}
\usepackage{tikz}
\usetikzlibrary{arrows}
\usetikzlibrary{backgrounds,fit}

\usepackage{amsmath}
\usepackage{amssymb}

\usepackage{hyperref}
\hypersetup{
    colorlinks=true,
    linkcolor=blue,
    filecolor=magenta,      
    urlcolor=cyan,
    pdftitle={Overleaf Example},
    pdfpagemode=FullScreen,
    }

\title{\textit{EfficientSpeech}: An On-Device Text to Speech Model}
\name{Rowel Atienza\thanks{Supported by Sibyl.AI to make AI accessible to everything and everyone.}}
\address{Electrical and Electronics Engineering Institute and AI Graduate Program, University of the Philippines
\\
rowel@eee.upd.edu.ph}
\begin{document}
%
\maketitle
\begin{abstract}
State of the art (SOTA) neural text to speech (TTS) models can generate natural-sounding synthetic voices. These models are characterized by large memory footprints and substantial number of operations due to the long-standing focus on speech quality with cloud inference in mind. Neural TTS models are generally not designed to perform standalone speech syntheses on resource-constrained and no Internet access edge devices. In this work, an efficient neural TTS called \textit{EfficientSpeech} that synthesizes speech on an ARM CPU in real-time is proposed. \textit{EfficientSpeech} uses a shallow non-autoregressive pyramid-structure transformer forming a U-Network. \textit{EfficientSpeech} has 266k parameters and consumes 90 MFLOPS only or about 1\% of the size and amount of computation in modern compact models such as \textit{Mixer-TTS}. \textit{EfficientSpeech} achieves an average mel generation real-time factor of 104.3  on an RPi4. Human evaluation shows only a slight degradation in audio quality as compared to \textit{FastSpeech2}.   
\end{abstract}

\tikzstyle{encoder}=[draw, fill=blue!20, minimum width=2cm, minimum height=0.5cm, minimum width=2.5cm]
\tikzstyle{encoder_box} = [draw=blue!50, thick, fill=blue!10, rounded corners, rectangle, inner sep=0.25cm, minimum height=3cm, ]

\tikzstyle{transformer_box} = [draw=blue!50, thick, fill=blue!10, rectangle, inner sep=0.25cm, minimum height=3cm, ]

\tikzstyle{up}=[draw, fill=red!20, minimum height=0.5cm, minimum width=2.5cm]
\tikzstyle{up_box} = [draw=red!50, thick, fill=red!10, rounded corners, rectangle, inner sep=0.25cm, minimum height=3cm, ]

\tikzstyle{conv_block}=[draw, fill=green!20, minimum height=0.5cm]

\tikzstyle{acoustic}=[draw, fill=green!20, minimum width=1.5cm, minimum height=0.5cm]
\tikzstyle{acoustic_box} = [draw=green!50, thick, fill=green!10, rounded corners, rectangle, inner sep=0.25cm, minimum height=3cm, ]

\tikzstyle{fuse}=[draw, fill=yellow!20, minimum width=2cm, minimum height=0.5cm]
\tikzstyle{fuse_box} = [draw=yellow!50, thick, fill=yellow!10, rounded corners, rectangle, inner sep=0.25cm, minimum height=3cm, ]

\tikzstyle{mel}=[draw, fill=orange!20, minimum width=2cm, minimum height=0.5cm, minimum width=2.5cm]
\tikzstyle{mel_box} = [draw=orange!50, thick, fill=orange!10, rounded corners, rectangle, inner sep=0.25cm, minimum height=3cm, ]

\tikzstyle{model}=[draw, fill=gray!20, minimum width=2cm, minimum height=0.5cm, minimum width=2.5cm]
    
\tikzstyle{image}=[minimum size=2cm]

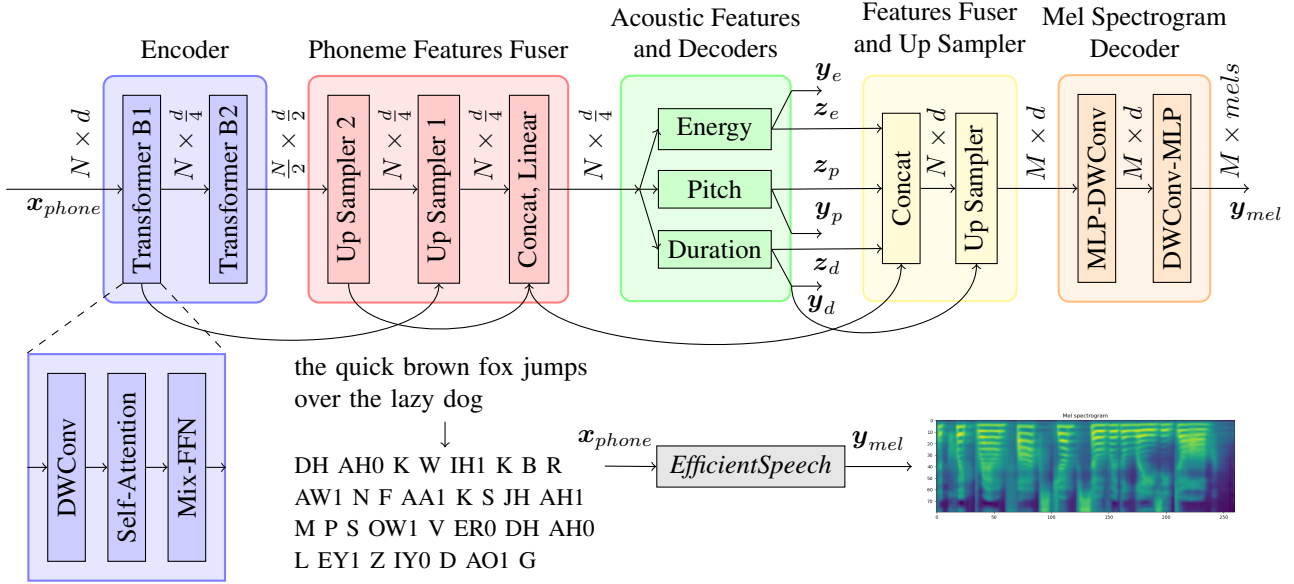
\begin{figure*}[t]
\begin{center}

\begin{tikzpicture} 
    \node (input_enc1) [coordinate] {};
    \node [encoder, right of=input_enc1, rotate=90, node distance=1.8cm](encoder1) {Transformer B1};
    \path[->](input_enc1)  edge node[below, midway] {$\boldsymbol{x}_{phone}$}  (encoder1.north);  
    
    \node [encoder, right of=encoder1, rotate=90, node distance=1.15cm](encoder2) {Transformer B2};
    \path[->](encoder1.south) edge node[midway, above, xshift=0.35cm, yshift=0.6cm, rotate=90] {$N\times\frac{d}{4}$} (encoder2.north);    
    
\begin{pgfonlayer}{background}
    \node[encoder_box] [fit = (encoder1) (encoder2)][label={[shift={(0,.1cm)}]Encoder}]{};
\end{pgfonlayer}

    \node [encoder, below of=encoder1, rotate=90, node distance=3.7cm, yshift=1cm](encoder_conv) {DWConv};
    \node [encoder, right of=encoder_conv, rotate=90, node distance=0.8cm](encoder_attention) {Self-Attention};
    \node [encoder, right of=encoder_attention, rotate=90, node distance=0.8cm](encoder_mixffn) {Mix-FFN};

\begin{pgfonlayer}{background}
    \node[transformer_box](transformer_box) [fit = (encoder_conv) (encoder_attention) (encoder_mixffn)][label={[shift={(0,.1cm)}]}]{};
\end{pgfonlayer}

    \path[-, dashed](encoder1.south west) edge node{}(transformer_box.north east);  
    \path[-, dashed](encoder1.north west) edge node{}(transformer_box.north west); 
    
    \path[->](encoder_conv.south) edge node{}(encoder_attention.north); 
    \path[->](encoder_attention.south) edge node{}(encoder_mixffn.north); 

    \path[->](transformer_box.west) edge node{}(encoder_conv.north); 
    \path[->](encoder_mixffn.south) edge node{}(transformer_box.east); 
    
    \node [up, right of=encoder2, rotate=90, node distance=1.6cm](up2) {Up Sampler 2};
  
    \node [up, right of=up2, rotate=90, node distance=1.2cm](up1) {Up Sampler 1};

    \node [up, right of=up1, rotate=90, node distance=1.2cm](concat1) {Concat, Linear};    
    
    \path[->](encoder2.south) edge node[midway, above, xshift=0.35cm, yshift=0.6cm, rotate=90] {$\frac{N}{2}\times\frac{d}{2}$} (up2.north);  

    \coordinate [xshift=0cm](mid_up2_up1) at (up1.north) ;
        
    \path[->](up2.south) edge node[midway, above, xshift=0.35cm, yshift=0.6cm, rotate=90] {$N\times\frac{d}{4}$} (mid_up2_up1); 
    \path[->](up1.south) edge node[midway, above, xshift=0.35cm, yshift=0.6cm, rotate=90] {$N\times\frac{d}{4}$} (concat1.north);

    \draw[->] (up2.west) .. controls +(down:8mm) and +(down:8mm) .. (concat1.west);
    
\begin{pgfonlayer}{background}
    \node[up_box] (up_box) [fit = (up1) (up2) (concat1)][label={[shift={(0,.1cm)}]Phoneme Features Fuser}]{};
\end{pgfonlayer}

    \draw[->] (encoder1.west) .. controls +(down:10mm) and +(down:10mm) .. (up1.west);

    \node (input_enc1_label)[rotate=90] at (1cm, 0.6cm) {$N\times{d}$}; 
    
    

    \coordinate [xshift=1.2cm](acoustic) at (concat1.south) ;
    
    \node [acoustic, right of=acoustic, node distance=1cm](pitch) {Pitch};
    \node [acoustic, above of=pitch, node distance=0.8cm](energy) {Energy};
    \node [acoustic, below of=pitch, node distance=0.8cm](duration) {Duration};    
    \path[->](concat1.south) edge node[midway, above, xshift=0.35cm, yshift=0.6cm, rotate=90] {$N\times\frac{d}{4}$} (acoustic.north);  
    \path[->](acoustic.south) edge node{} (energy.west);  
    \path[->](acoustic.south) edge node{} (pitch.west);
    \path[->](acoustic.south) edge node{} (duration.west);

\begin{pgfonlayer}{background}
    \node[acoustic_box](acoustic_box) [fit = (acoustic) (pitch) (energy) (duration) ][label={[shift={(0,.1cm)},align=center]Acoustic Features \\ and Decoders}]{};
\end{pgfonlayer}

    \coordinate [yshift=1.3cm](energy_mid) at (acoustic_box.east) ;
    \coordinate [xshift=0.4cm](energy_out) at (energy_mid);
    \path[-](energy.east)  edge node  {} (energy_mid);
    \path[->](energy_mid)  edge  node[above, midway, xshift=0.3cm] {$\boldsymbol{y}_{e}$} (energy_out);
 
    \coordinate [yshift=-0.6cm](pitch_mid) at (acoustic_box.east);
    \coordinate [xshift=0.4cm](pitch_out) at (pitch_mid);
    \path[-](pitch.east)  edge node  {} (pitch_mid);
    \path[->](pitch_mid)  edge  node[above, midway, xshift=0.3cm] {$\boldsymbol{y}_{p}$} (pitch_out); 
    
    \coordinate [yshift=-1.3cm](duration_mid) at (acoustic_box.east);
    \coordinate [xshift=0.4cm](duration_out) at (duration_mid);
    \path[-](duration.east)  edge node  {} (duration_mid);
    \path[->](duration_mid)  edge  node[below, midway, xshift=0.2cm] {$\boldsymbol{y}_{d}$} (duration_out);

    \node [fuse, right of=acoustic_box, rotate=90, node distance=2.6cm](concat2) {Concat};
    
    \node [fuse, right of=concat2, rotate=90, node distance=1cm](up3) {Up Sampler};

    \coordinate [yshift=0.8cm](concat2_northeast) at (concat2.north);
    \coordinate [yshift=-0.8cm](concat2_northwest) at (concat2.north);
    
    \draw[->] (concat1.west) .. controls +(down:10mm) and +(down:15mm) .. (concat2.west);
    
    \draw[->] (duration_mid) .. controls +(down:10mm) and +(down:10mm) .. (up3.west);
    
    \path[->](concat2.south) edge node[midway, above, xshift=0.25cm, yshift=0.6cm, rotate=90] {$N\times{d}$} (up3.north);
    \path[->](pitch.east) edge node[above, midway] {$\boldsymbol{z}_{p}$} (concat2.north);
    \path[->](energy.east) edge node[above, midway] {$\boldsymbol{z}_{e}$} (concat2_northeast);
    \path[->](duration.east) edge node[below, midway] {$\boldsymbol{z}_{d}$} (concat2_northwest);
    
\begin{pgfonlayer}{background}
    \node[fuse_box] [fit = (concat2) (up3) ][label={[shift={(0,.1cm)},align=center]Features Fuser \\ and Up Sampler}]{};
\end{pgfonlayer}

    
    \node [mel, right of=up3, rotate=90, node distance=1.6cm](mel_mlp) {MLP-DWConv};
    \node [mel, right of=mel_mlp, rotate=90, node distance=1cm](mel_block1) {DWConv-MLP};

    \path[->](up3.south) edge node [midway, above, xshift=0.25cm, yshift=0.6cm, rotate=90] {$M\times{d}$} (mel_mlp.north);
    \path[->](mel_mlp.south) edge node[midway, above, xshift=0.25cm, yshift=0.6cm, rotate=90] {$M\times{d}$} (mel_block1.north);
    
\begin{pgfonlayer}{background}
    \node[mel_box](mel_box) [fit = (mel_mlp) (mel_block1)  ][label={[shift={(0,.1cm)},align=center]Mel Spectrogram \\ Decoder}]{};
\end{pgfonlayer}
    
    \coordinate [xshift=0.5cm](mel_out) at (mel_box.east);
   
    \path[->](mel_block1.south) edge node[midway, above, xshift=0.4cm, yshift=0.9cm, rotate=90] {$M\times{mels}$} (mel_out);

    \node [xshift=0.1cm, yshift=-0.3cm] (mel_out_label) at (mel_out) {$\boldsymbol{y}_{mel}$} ;  

    \coordinate [yshift=-2.6cm,xshift=1.cm](input_text) at (up2.south);
    \node[text width=4cm, align=left](x_text) at (input_text) {the quick brown fox jumps over the lazy dog};
    \coordinate [yshift=-4.3cm,xshift=1.cm](input_phoneme) at (up2.south);
    \node[text width=4cm, align=left](x_phoneme) at (input_phoneme) {\small DH AH0 K W IH1 K B R AW1 N F AA1 K S JH AH1 M P S OW1 V ER0 DH AH0 L EY1 Z IY0 D AO1 G};
    \path[->](x_text.south) edge  (x_phoneme.north);

    \node [image] (image) [below of=mel_mlp,node distance=3.7cm, xshift=-0.2cm] {\includegraphics[width=0.24\textwidth]{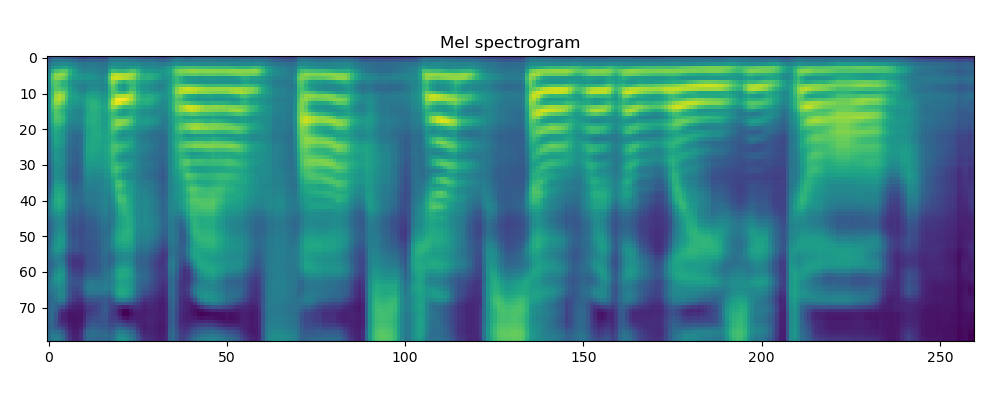}};

    \node [model, left of=image, node distance=4.4cm](efficientspeech) {\textit{EfficientSpeech}};
    \path[->](efficientspeech.east) edge node[midway, above, xshift=0cm, yshift=0.1cm] {$\boldsymbol{y}_{mel}$} (image.west); 

    \coordinate [yshift=-0.25cm](input_phoneme_coord) at (x_phoneme.north east);
    
    \path[->](input_phoneme_coord) edge node[midway, above, xshift=-0.2cm, yshift=0.1cm] {$\boldsymbol{x}_{phone}$} (efficientspeech.west);

\end{tikzpicture}

\end{center}
   \caption{Model architecture of \textit{EfficientSpeech}. The phoneme encoder is made of two transformer encoder blocks fused with up sampled features resembling a U-Net. \textit{EfficientSpeech} uses parallel acoustic features and outputs prediction. Acoustic features are merged with phoneme features and up sampled for mel-spectrogram decoding which is made of two blocks.}
\label{fig:model_architecture}
\end{figure*}

\section{Introduction}
\label{sec:intro}

Voice is one of our primary means of communication. If our devices can also speak, a new type of natural interaction with electronic gadgets and appliances is feasible. Even better, if devices can perform standalone voice synthesis without relying on cloud services, new applications and advantages will emerge. For instance, a WiFi router can tell us what went wrong when there is no Internet access. A smart camera installed in a remote area can warn intruders. These useful actions are done by the device in autonomous manner and without relying on cloud services. As added benefits of on-device voice synthesis, privacy issues are mitigated, robustness is enhanced, and high responsiveness, low-latency and availability can be guaranteed. 

In terms of natural sounding voice generation, neural TTS systems such as \textit{FastSpeech2} \cite{DBLP:conf/iclr/0006H0QZZL21}, \textit{FastPitch} \cite{lancucki2021fastpitch}, \textit{Tacotron2} \cite{shen2018natural}, \textit{Deep Voice 3} \cite{ping2018deep}, \textit{TransformerTTS} \cite{li2019neural} and \textit{Mixer-TTS} \cite{tatanov2022mixer} dominate the state of the art performance in MOS scores. These neural TTS models are designed with AI accelerators such as GPUs or TPUs in mind. There is little emphasis on investigating the feasibility of achieving standalone on-device model inference. In particular, autoregressive models like \textit{Tacotron2}, \textit{Deep Voice 3} and \textit{TransformerTTS} are inherently slow. While non-autoregressive neural TTS such as \textit{FastSpeech2} and \textit{Mixer-TTS} are fast and have competitive voice quality that is comparable to autoregressive counterparts, these models have big footprints making them unsuitable for memory-constrained edge devices.  

Recent attempts to build on-device neural TTS include \textit{On-device TTS} \cite{achanta2021device}, \textit{LiteTTS} \cite{nguyen2021litetts}, \textit{PortaSpeech} \cite{ren2021portaspeech}, \textit{LightSpeech} \cite{luo2021lightspeech} and \textit{Nix-TTS} \cite{chevi2022nix}.   \textit{On-device TTS} is slow and resource intensive since it is a modified \textit{Tacotron2} for mel spectrogram generation and uses \textit{WaveRNN} for vocoder. Though \textit{LiteTTS} can generate voice from text, it is still resource intensive with 13.4M parameters. In addition, two-stage TTS models are still better in terms of both training stability and synthetic voice quality. \textit{PortaSpeech} uses VAE and Flow models to generate mel spectrogram. The smallest version has 6.7M parameters and is  characterized by noticeable voice quality deterioration. \textit{LightSpeech} uses neural architecture search (NAS) to reduce the model size of \textit{FastSpeech2}. While the resulting model is small at 1.8M parameters, the NAS process is notoriously compute intensive with a huge environmental impact. Furthermore, NAS is susceptible to overfitting. A model architecture optimized on one language dataset (e.g.\ English) is not guaranteed to work on another (e.g.\ Korean). \textit{Nix-TTS} applied knowledge distillation to reduce the size of \textit{VITS} \cite{kim2021conditional} to 5.2MB by separately training text-to-latent encoder and latent-to-waveform decoder. While there is a significant reduction in size, the decoder is single-use or encoder specific unlike general purpose vocoders such as \textit{HiFiGAN} \cite{kong2020hifi} that is available in a sub-1M-parameter model for edge devices. Ironically, while the above mentioned  models promote on-device TTS, there was no validation done on ARM CPUs except for \textit{Nix-TTS} that used a compiled ONNX model. Furthermore, most of these models have no publicly available implementations. Thus, reproducibility, fair comparison and analysis are difficult to perform. 

In this paper, \textit{EfficientSpeech}, a natural sounding TTS model that is suitable for edge devices is proposed. \textit{EfficientSpeech} is using a shallow U-Network \cite{ronneberger2015u} pyramid transformer phoneme encoder and a shallow transposed convolutional block as the mel spectrogram decoder. \textit{EfficientSpeech} has 266k parameters only, about 15\% of the size of \textit{LightSpeech} or 0.8\% of \textit{FastSpeech2}. \textit{EfficientSpeech} consumes 90 MFLOPS only to generate 6 sec of mel spectrogram. Using the compact version of HiFiGAN \cite{kong2020hifi}, the total model parameters is 1.2M or 22\% of text to speech waveform \textit{Nix-TTS}. Using HiFiGAN as vocoder, it runs at an RTF of 1.7 for voice generation on RPi4. Without the vocoder overhead, the mel spectrogram generation is at RTF speed of 104.3. \textit{EfficientSpeech} achieves a competitive CMOS of -0.14 when trained on LJSpeech dataset \cite{ljspeech17} and evaluated against \textit{FastSpeech2}. Due to its small size, \textit{EfficientSpeech} can be trained on a single GPU in 12hrs.

\section{Model Architecture}
\label{sec:majhead}
Figure \ref{fig:model_architecture} shows the model architecture of \textit{EfficientSpeech}. The phoneme sequence $\boldsymbol{x}_{phone}\in\mathbb{R}^{N\times{d}}$ is an embedding of the input text phonemes. All convolutional layers are 1D. $N$ is the variable phoneme sequence length while $d=128$ is the embedding size.

The \textit{Phoneme Encoder} is made of 2 transformer blocks. Each block is made of a depth-wise separable convolution for feature merging, \textit{Self-Attention} between merged features and \textit{Mix-FFN} for non-linear feature extraction. \textit{Mix-FFN} is similar to a typical transformer \cite{vaswani2017attention}  \textit{FFN} except for an additional convolution layer and the use of GeLU \cite{hendrycks2016gaussian} activation between two linear layers. Layer Normalization (\textit{LN}) \cite{ba2016layer} is applied after \textit{Self-Attention} and \textit{Mix-FFN}. Both \textit{Self-Attention} and \textit{Mix-FFN} use residual connection for fast convergence.

The first transformer block retains the sequence length while reducing the feature dimension by $\frac{1}{4}$. The second transformer block reduces the sequence length by half while doubling the feature dimension. Each transformer block output feature is upsampled using a linear layer and a transposed convolutional layer. An identity layer replaces the transposed convolution if the target feature shape of $N\times{\frac{d}{4}}$ is already in place. Both features are then fused together to form the final phoneme features. This U-Network \cite{ronneberger2015u} style of architecture was inspired by \textit{SegFormer} \cite{xie2021segformer} for semantic segmentation in computer vision. Reducing the feature dimension and sequence length lowers the FLOPS and the number of parameters of the model.  

The \textit{Acoustic Features and Decoders} block borrows the idea from \textit{Variance Adaptor} of \textit{FastSpeech2}. It forces the network to predict the \textit{Energy}: $\boldsymbol{y}_e$, \textit{Pitch}: $\boldsymbol{y}_p$ and \textit{Duration}: $\boldsymbol{y}_d$. The difference in our implementation is that instead of predicting the acoustic parameters in series, \textit{EfficientSpeech} generates them in parallel which results to a faster inference. The predicted values of, \textit{Energy}: $\boldsymbol{y}_e$, \textit{Pitch}: $\boldsymbol{y}_p$ and \textit{Duration}: $\boldsymbol{y}_d$, are generated by 2 blocks of \textit{Conv-LN-ReLU} and a final linear layer (with \textit{ReLU} for duration to ensure positive values).  The binned energy and pitch features are embedded at the last layer to produce \textit{Energy}: $\boldsymbol{z}_e$ and \textit{Pitch}: $\boldsymbol{z}_p$. Meanwhile, \textit{Duration}: $\boldsymbol{z}_d$ is extracted before the \textit{ReLU} activation.

At the \textit{Features Fuser and Up Sampler} block, all acoustic features are reused and fused together with the phoneme features. The fused features are then up sampled to the correct mel sequence length $M$ using the predicted \textit{Duration}: $\boldsymbol{y}_{d}$.

The last stage is the \textit{Mel Spectrogram Decoder}. It is made of 2 blocks of a linear layer and two layers of depth-wise separable convolution. Each layer uses \textit{Tanh} activation followed by \textit{LN}.

\subsection{Model Training}
The dataset used for training is LJSpeech \cite{ljspeech17} that is made of $13,100$ audio clips with corresponding text transcripts. $12,588$ samples are set aside for training while $512$ clips are for testing. The phoneme sequence is generated using \textit{g2p} \cite{g2pE2019}, an open-source English grapheme (spelling) to phoneme (pronunciation) converter. The waveform is transformed into mel spectrogram with window and FFT lengths of $1,024$, hop length of $256$ and sampling rate of $22,050$. The resulting mel spectrogram has $80$ channels.

Montreal Force Alignment (MFA) \cite{mcauliffe2017montreal} is used to establish the target phoneme duration. Pitch and energy ground truth values are computed using STFT and WORLD vocoder \cite{morise2009fast} respectively.

The total loss function is shown in Equation \ref{eq:loss_function}. Mel spectrogram loss function $\mathcal{L}_{mel}$ is \textit{L1} with $\alpha=10$. $MSE$ is used for \textit{Pitch}: $\mathcal{L}_{p}$, \textit{Energy}: $\mathcal{L}_{e}$, and \textit{Duration}: $\mathcal{L}_{d}$ loss functions. $\beta=2$, $\gamma=2$ and $\lambda=1$. 

\begin{align}\label{eq:loss_function}
\mathcal{L} = \alpha\mathcal{L}_{mel} + \beta\mathcal{L}_{p} + \gamma\mathcal{L}_{e} + \lambda\mathcal{L}_{d}.
\end{align}

The \textit{EfficientSpeech} model is trained for $5,000$ epochs. Batch size is $128$. The optimizer is AdamW \cite{loshchilov2018decoupled} with learning rate of $0.001$, cosine learning rate decay and warm up of $50$ epochs.

\begin{table}[]
    \centering
   
    \begin{tabular}{l |r | r }
     \hline
     {} & \# Parameters &  ES Relative  \\
     Model & (M)$\downarrow$ & \#  Parameters  \\
     \hline
    \textit{EfficientSpeech} (ES) & $\textbf{0.27}$ & $-$ \\
    \textit{FastSpeech2}\cite{DBLP:conf/iclr/0006H0QZZL21} & $30.81$ & $0.86\%$ \\
    \textit{Tacotron2}\cite{shen2018natural} & $23.81$ & $1.12\%$ \\
    \textit{MixerTTS}\cite{tatanov2022mixer} & $20.06$ & $1.33\%$ \\
    \textit{LightSpeech}\cite{luo2021lightspeech} & $1.80$ & $14.78\%$ \\
    
     \hline
    \end{tabular}
    \caption{The number of parameters in different mel spectrogram generator models. \textit{LightSpeech} is based on published data.}
    \label{tab:params}
\end{table}

\begin{table}[]
    \centering
   
    \begin{tabular}{l |r | r }
     \hline
     {} & {} & ES Relative   \\
     Model & GFLOPS $\downarrow$ & GFLOPS  \\
     \hline
    \textit{EfficientSpeech} (ES) & $\textbf{0.09}$ & $-$ \\
    \textit{FastSpeech2}\cite{DBLP:conf/iclr/0006H0QZZL21} & $15.87$ & $0.57\%$ \\
    \textit{Tacotron2}\cite{shen2018natural} & $16.20$ & $0.56\%$ \\
    \textit{MixerTTS}\cite{tatanov2022mixer} & $10.29$ & $0.87\%$ \\
    \textit{LightSpeech}\cite{luo2021lightspeech} & $0.76$ & $11.84\%$ \\
    
     \hline
    \end{tabular}
    \caption{Amount of computations in terms of GFLOPS in different mel spectrogram generator models. Average voice length is 6 sec. \textit{LightSpeech} is based on published data for 9 sec of speech.}
    \label{tab:flops}
\end{table}

\begin{table*}[]
    \centering
   
    \begin{tabular}{l |r | r | r | r | r | r }
     \hline
     {} & mRTF & ES Relative & mRTF & ES Relative & mRTF & ES Relative  \\
     Model & V100 $\uparrow$& Speed-up & Xeon 2.2G $\uparrow$& Speed-up & ARM 1.5G $\uparrow$ & Speed-up \\
     \hline
    \textit{EfficientSpeech} (ES) & $\textbf{953.3}$  & $-$ & $\textbf{470.2}$ & $-$ & $\textbf{104.3}$ & $-$ \\
    \textit{FastSpeech2}\cite{DBLP:conf/iclr/0006H0QZZL21} & $371.3$ & $2.6\times$ & $64.7$ & $7.3\times$ & $5.2$ & $20.1\times$ \\
    \textit{Tacotron2}\cite{shen2018natural} & $8.3$ & $114.7\times$ & $1.2$ & $379.4\times$ & $0.2$ & $462.2\times$ \\
    \textit{MixerTTS}\cite{tatanov2022mixer} & $204.9$ & $4.7\times$ & $55.2$ & $8.5\times$ & $2.9$ & $36.5\times$\\
    \textit{LightSpeech}\cite{luo2021lightspeech} & $-$ & $-$ & $107.5$ & $4.4\times$ & $-$ & $-$ \\
    
     \hline
    \end{tabular}
    \caption{mRTF is the average of number of seconds of speech divided by the mel generation time for 128 samples from the test split. \textit{LightSpeech} is from published data on Xeon 2.6GHz and it was not tested on other processors. The benchmarks were done on NVIDIA V100 32GB, Intel Xeon CPU E5-2650 v4 @ 2.20GHz and Raspberry Pi 4 Model B BCM2711 Quad Cortex A72 (ARMv8) 64-bit 1.5GHz.}
    \label{tab:mel-rtf}
\end{table*}

\begin{table*}[]
    \centering
   
    \begin{tabular}{l |r | r | r | r | r | r }
     \hline
     {} & RTF  & ES Relative & RTF & ES Relative & RTF & ES Relative  \\
     Model & V100 $\uparrow$ & Speed-up &  Xeon 2.2G $\uparrow$& Speed-up & ARM 1.5G $\uparrow$& Speed-up \\
     \hline
    \textit{EfficientSpeech} (ES) & $\textbf{363.0}$ & $-$ & $\textbf{24.1}$ & $-$ & $\textbf{1.7}$ & $-$ \\
    \textit{FastSpeech2}\cite{DBLP:conf/iclr/0006H0QZZL21} & $66.9$ & $5.4\times$ & $11.9$ & $2.0\times$ & $1.3$ & $1.3\times$ \\
    \textit{Tacotron2}\cite{shen2018natural} & $7.7$ & $47.3\times$ & $1.0$ & $24.9\times$ & $0.1$ & $12.4\times$ \\
    \textit{MixerTTS}\cite{tatanov2022mixer} & $56.6$ & $6.4\times$ & $6.4$ & $3.8\times$ & $0.2$ & $6.9\times$\\
    
     \hline
    \end{tabular}
    \caption{RTF is the average of number of seconds of speech divided by the waveform generation time for 128 samples from the test split. See Table \ref{tab:mel-rtf} on the hardware specifications. No available data for \textit{LightSpeech}. }
    \label{tab:wav-rtf}
\end{table*}

\begin{table}[]
    \centering
   
    \begin{tabular}{l |r  }
     \hline
     Model & CMOS$\uparrow$ \\
     \hline
    \textit{FastSpeech2}\cite{DBLP:conf/iclr/0006H0QZZL21} & $0.0$  \\
    \textit{EfficientSpeech} & $-0.14$  \\
    \textit{LightSpeech}\cite{luo2021lightspeech} & $0.04$ \\
    
     \hline
    \end{tabular}
    \caption{The CMOS between \textit{FastSpeech2} and \textit{EfficientSpeech}. For reference, we include the published results of \textit{LightSpeech}.}
    \label{tab:mos}
\end{table}

\section{Experimental Results}
\label{ssec:subhead}

The \textit{EfficientSpeech} evaluation is  not only in terms of the generated speech quality but also its trade off with respect to the number of parameters, amount of computations as measured by floating point operations (FLOPS), and speed or throughput in terms of latency. A comprehensive benchmark enables us to get the overall picture of our model performance as a function of memory, computational budget and time \cite{dehghani2021efficiency} instead of focusing only on selected favorable metrics. 

The number of parameters is commonly used as a proxy to the amount of memory needed by the model during execution.  FLOPS reflects the number of Fused-Multiply-Add (FMA) operations needed to complete an inference. For variable input text sequence length like in TTS, FLOPS is measured using 128 randomly sampled text inputs from the test split. FLOPS increases with input text length. Latency is measured in terms of the number of seconds of voice generated per second or the real-time-factor (RTF). The inverse of this RTF, the time needed to generate 1 sec of voice, can also be used but it leads to small fractional numbers that are less intuitive to interpret. To focus on the speed of \textit{EfficientSpeech}, mel spectrogram real-time-factor (mRTF) is introduced. mRTF is the number of seconds of speech divided by the mel generation time.

\texttt{fvcore} \cite{fvcore} is used to compute the number of parameters and FLOPS. Time measurements use the CPU wall clock. Table \ref{tab:params} shows the number of parameters and the relative footprint of \textit{EfficientSpeech} in comparison with state-of-the-art mel spectrogram generators. \textit{EfficientSpeech} is tiny at 266k parameters leading to a very small number of FLOPS as shown in Table \ref{tab:flops}. The effect of the small number of parameters and FLOPS is a fast mel spectrogram generation reaching mRTF of 953.3 on a V100 GPU as shown in Table \ref{tab:mel-rtf}. The speed is more evident on an RPi4 ARM CPU where \textit{EfficientSpeech} reaches mRTF of 104.3 which is $20.1\times$ faster compared to \textit{FastSpeech2}.

For \textit{Tacotron2} and \textit{MixerTTS}, the pre-trained versions provided by NVIDIA NeMo \cite{kuchaiev2019nemo} with HiFiGANv1 was evaluated. For speech generation, both models are unable to run with RTF $\ge$ 1.0 on the ARM CPU of RPi4. Furthermore, NeMo employed mixed precision training and other optimizations providing a significant acceleration in GPUs.

Table \ref{tab:mos} shows the CMOS \cite{loizou2011speech} as evaluated by 15 participants with high English listening comprehension. The synthesized speech waveforms are from the test split. Both \textit{EfficientSpeech} and \textit{FastSpeech2} used the small version of off-the-shelf HiFiGANv2 with 0.9M parameters. In terms of audio quality, \textit{EfficientSpeech} outputs only suffer a slight degradation in quality in spite of its small size. For reference, the published CMOS score of \textit{LightSpeech} as compared to \textit{FastSpeech2} is also shown. However, note that the samples used to obtain this score are not available.


\section{Discussion}
The RTF slow down from Table \ref{tab:mel-rtf} to \ref{tab:wav-rtf}, can be attributed to the inefficient vocoder. At mRTF of 104.3 on RPi4, \textit{EfficientSpeech} has a significant headroom to speed up the voice generation given a counter part lightweight vocoder. In the experimental setup, the HiFiGAN consumes 5.0 GFLOPS while the \textit{EfficientSpeech} model overhead is only 0.09 GFLOPS. Meanwhile, majority of SOTA mel generator models have used up most of RPi4 Model B 13.5 to 32 GFLOPS (estimates vary).

The computational performance of low-cost BCM2835 SoC ARMv6 256MB to 512MB RAM used in RPi Zero, A and B is about 0.2 to 0.3 GFLOPS giving \textit{EfficientSpeech} enough leeway but not for the vocoder. RPi3 Model B BCM2837/B0 SoC ARMv7/8 1GB RAM has a computing performance of about 3.6 to 6.2 GFLOPS. RPi2 Model B BCM2836 and  BCM2837 SoCs ARMv7 1GB RAM has about 1.5 to 4.4 GFLOPS. Theoretically, a sub 0.1 GFLOPS vocoder will enable wide adoption of neural TTS such as \textit{EfficientSpeech} on many low-cost and low-power devices. A sub 1 GFLOPS vocoder can already broaden the device coverage of neural TTS to RPi2.  At 266k parameters, 16-bit floating point, the footprint of \textit{EfficientSpeech} is about 532kb leaving enough RAM space to store results of intermediate layers even on low memory 256MB SoCs. 

Note that although the number of model parameters and FLOPS have impact on RTF, there are other factors that may contribute to latency. For instance, a model architecture that has dense skip connections has inherent delays in the forward propagation due to buffering. Models with many layers are slow due to the increasing forward propagation steps. Feature dimensions mismatch, normalization layers and complex activation functions can also cause slow model inference.


\section{Conclusion}
 The quality voice synthesis improves as the model size   increases.  \textit{EfficientSpeech} code and pre-trained weights are available on GitHub for:   Tiny (266k), Small (952k) and Base (4M). See: \href{https://github.com/roatienza/efficientspeech}{https://github.com/roatienza/efficientspeech}

\section{Acknowledgement}
Project funding by Rowel Atienza through Sibyl.AI. Conference attendance funding by ERDT-FRDG.



\bibliographystyle{IEEEbib}
\bibliography{refs}

\end{document}